\documentclass[structabstract]{aa}
\usepackage{graphicx}
\usepackage{txfonts}
\begin{document}
\title{Non-linear modelling of beat Cepheids: Resonant and non-resonant models}

   \author{R. Smolec\inst{1}
          \and
           P. Moskalik\inst{2}
          }
   \institute{Institute for Astronomy (IfA), University of Vienna,
              T\"urkenschanzstrasse 17, A-1180 Vienna, Austria\\
              \email{radek.smolec@univie.ac.at}
         \and
             Copernicus Astronomical Centre, Bartycka 18, 00-716 Warszawa, Poland\\
             \email{pam@camk.edu.pl}
             }

   \date{Received ; accepted}

% \abstract{}{}{}{}{}
% 5 {} token are mandatory

  \abstract
{Double-periodic (beat) Cepheids are important astrophysical objects
which allow testing both the stellar evolution and stellar
pulsation theories, as well as the physical properties of matter in
stellar conditions. However, the phenomenon of double-periodic
pulsation is still poorly understood. Recently we
rediscussed the problem of modelling the double-periodic pulsation with
non-linear hydrocodes. We showed that the published non-resonant
double-mode models are incorrect, because they exclude the
negative buoyancy effects.}
   {We continue our efforts to verify whether the Kuhfu\ss{}
one-equation convection model with negative buoyancy
included can reproduce the double-periodic Cepheid
pulsation.}
   {Using the direct time integration hydrocode, which
implements the Kuhfu\ss{} convection model, we search for stable
double-periodic Cepheid models. We search for models pulsating in
both fundamental and first overtone modes (F+1O), as well as in
the two lowest order overtones (1O+2O). In the latter case,
we focus on reproducing double-overtone Cepheids of the
Large Magellanic Cloud (LMC).}
   {We have found full amplitude non-linear beat Cepheid models of
both types, F+1O and 1O+2O. In the case of F+1O models, the
beat pulsation is most likely caused by the three-mode resonance,
$2\omega_1=\omega_0+\omega_2$, while in the double-overtone
models the underlying mechanism (resonant or non-resonant) cannot be
identified beyond doubt. Double-periodic models found in our survey
exist, however, only in narrow period ranges and cannot explain
the majority of the observed double-periodic objects.}
   {With only little doubt left, we conclude that current one
dimensional one-equation convection models are incapable of
reproducing the majority of the observed beat Cepheids. Among
the shortcomings of current pulsation hydrocodes, the simple treatment
of convection seems to be the most severe one. Growing
evidence for the presence of non-radial modes in Cepheids suggests
that the interaction between radial and non-radial modes should
also be investigated.}


   \authorrunning{R. Smolec \& P. Moskalik}
   \titlerunning{Non-linear modelling of beat Cepheids}

   \maketitle

%-_-_-_-_-_-_-_-_-_-_-_-_-_-_-_-_-_-_-_-_-_-_-_-_-_-_-_-_-_-_-_-_-_-_-_-_-_
%-_-_-_-_-_-_-_-_-_-_-_-_-_-_-_-_-_-_-_-_-_-_-_-_-_-_-_-_-_-_-_-_-_-_-_-_-_

\section{Introduction}

Cepheids are periodic variable stars, pulsating radially with
periods ranging from a single day to over one hundred days for the most
luminous variables. They obey an empirical period-luminosity
relation, which ranks them among the most important distance
indicators in astrophysics. They populate a rather narrow strip in
the HR diagram, called the classical instability strip, in which
pulsation is driven by the opacity mechanism acting in
hydrogen-helium partial ionisation region. They are giant,
relatively cool stars with typical effective temperatures ranging
from $\log T_{\rm eff}=3.7$ to $\log T_{\rm eff}=3.8$ (see e.g.
Sandage, Tammann \& Reindl \cite{str04}). Estimated values of the
width of the instability strip do not exceed
$1000$\thinspace{}Kelvins.\footnote{The transformation from the 
observed colours to effective temperatures is particularly
difficult for pulsating stars. In addition, inaccurate or unknown
reddenings for individual objects significantly add to the uncertainty
of the $\log T_{\rm eff}$ calibration.} Depending on the mass, the star can
cross the instability region up to three times. The first crossing
occurs in the post-main sequence evolution, before helium ignites,
while the star evolves very quickly towards the red giant branch.
After the helium ignition, more massive stars enter the horizontal blue
loop, crossing the instability strip twice, first during the blue-ward and
then the red-ward evolution. The corresponding second and third crossings
last roughly two orders of magnitude longer than the first crossing.
Hence, most of the observed Cepheids are expected to be helium-burning 
objects.

Most of the Cepheids are single-periodic variables pulsating either
in the fundamental mode or in the first overtone mode. Only a few
examples of second overtone Cepheids are known. In many Cepheids
simultaneous pulsations in two modes are observed. Double-periodic
Cepheids (or beat Cepheids) are very interesting and important
astrophysical objects. Most of these variables pulsate
simultaneously in two consecutive radial pulsation modes, either in
the fundamental and in the first overtone (F+1O) or in the two
lowest order overtone modes (1O+2O, double-overtone Cepheids in the
following). The two pulsation periods can be used to constrain the
stellar parameters; in particular, the mass can be derived using the
Petersen diagram (Petersen \cite{pet73}). The disagreement between these
pulsation masses of F+1O beat Cepheids and the masses derived from
evolutionary computations was one of the factors that motivated the
revision of opacity tables (Simon \cite{sim82}). Multi-mode Cepheids
still provide useful and stringent tests for stellar evolution and
pulsation theories (see e.g. Moskalik \& Dziembowski \cite{md05},
Dziembowski \& Smolec \cite{ds09a}).

Despite the great importance of the beat Cepheids and the apparent
simplicity of their oscillation, the phenomenon of double-periodic
pulsation is not well understood. Many efforts were made over the
past forty years to study the interaction of pulsation modes and the
origin of the double-periodic pulsation. These include 
theoretical studies as well as numerical modelling.

The necessary condition for the double-periodic pulsation to occur
is simultaneous linear instability of the respective two modes. This
condition is not sufficient because mode selection
is a non-linear phenomenon. The interaction between pulsation modes
may be either of a non-resonant or of a
resonant character. In most of the observed Cepheids both
mechanisms lead to stable single-periodic pulsation. In the
non-resonant case, it is also a single-mode pulsation. The
distinction between multi-periodic and multi-mode pulsation is
important. In the resonant case the single-periodic
pulsation is actually multi-mode, because the resonantly coupled
mode is also excited to high amplitude. Owing to the non-linear
frequency synchronisation, it does not appear with a separate
frequency, but its presence is manifested in the distortion of light
and radial velocity curves. Bump Cepheids, which are
single-periodic, double-mode pulsators, provide an excellent example.
In this paper we deal with double-periodic pulsators and will use the 
term double-mode below only for stars and
models for which the resonant mechanism is excluded.

Double-periodic Cepheids are rare compared to single-periodic
Cepheids, and the identification of the conditions that lead to stable
double-periodic pulsation was and still is of key importance. Simon
(\cite{sim79}) suggested that the double-periodic pulsation may be a
resonant phenomenon, connected with the three-mode resonance,
$\omega_1+\omega_0=\omega_3$. Using the amplitude equation formalism
Dziembowski \& Kov\'acs (\cite{dk84}) showed that the resonance
proposed by Simon actually stabilizes the single-periodic pulsation.
They also pointed out that the 2:1 resonance between the linearly
excited mode and the linearly damped, parasite mode can lead to
double-periodic pulsation. This was later confirmed with radiative
RR~Lyrae and Cepheid models (Kov\'acs \& Buchler \cite{kb88},
Buchler, Moskalik \& Kov\'acs \cite{bmk90}, Smolec \cite{rs09a}). 
The models agreed not once with the observations, though. Indeed, 
a comparison of the observed periods and the period ratios of beat
Cepheids with the results of linear modelling indicate that in most of
the double-periodic Cepheids resonances of a low order are excluded,
and therefore the non-resonant mechanism should be 
operational. For many years radiative models failed to reproduce the
non-resonant double-mode pulsation in both RR~Lyrae stars and in
Cepheids. It was the inclusion of turbulent convection into the
hydrocodes that led to success. Koll\'ath et al. (\cite{koea98})
published the double-mode Cepheid models and Feuchtinger
(\cite{fe98}) published a double-mode RR~Lyrae model. Later, only
the Florida-Budapest group computed the surveys of double-mode
pulsation (Koll\'ath et al. \cite{koea02}, Szab\'o, Koll\'ath \&
Buchler \cite{skb04}, Buchler \cite{bu09}). The reliability of these
models was recently questioned however by the analysis of Smolec \&
Moskalik (\cite{sm08a,sm08b}) (see also
Sect.~\ref{M.negativebuoyancy}). We have shown that the computed
double-mode pulsation arises from the neglect of negative buoyancy
in the Florida-Budapest hydrocode, which is physically not
justified. Yet we could not find any non-resonant F+10 Cepheid models even with 
our pulsation hydrocode (Smolec \& Moskalik \cite{sm08a}, 
Sect.~\ref{M.code}), which properly included the negative buoyancy 
(Smolec \& Moskalik \cite{sm08b}).

In the present paper we discuss some {\it resonant} F+1O Cepheid
models that we computed (Sect.~\ref{R.par}). Although these
models are restricted to a very narrow parameter range, they agree
sufficiently well with the observations. Finally we present 
a survey of models which we conducted
in search for stable double-overtone (1O+2O) Cepheid pulsation
(Sect.~\ref{R.LMC}). This survey is complementary to our earlier
work, in which we searched for non-resonant F+1O Cepheid models 
(Smolec \& Moskalik \cite{sm08b}), and allows us to draw more definite 
conclusions about the ability of
current convection models to reproduce the beat Cepheid
pulsation (Sect.~\ref{C}). Preliminary results of this work were
presented in Smolec \& Moskalik (\cite{sm09}).

%-_-_-_-_-_-_-_-_-_-_-_-_-_-_-_-_-_-_-_-_-_-_-_-_-_-_-_-_-_-_-_-_-_-_-_-_-_

\section{Turbulent convection model and numerical methods}

%-_-_-_-_-_-_-_-_-_-_-_-_-_-_-_-_-_-_-_-_-_-_-_-_-_-_-_-_-_-_-_-_-_-_-_-_-_

\subsection{Turbulent convection model and pulsation hydrocodes\label{M.code}}

In all our computations we use the pulsation codes described by Smolec \& Moskalik (\cite{sm08a}). These are a static model builder, linear non-adiabatic code and a direct time-integration non-linear hydrocode. The codes use a simple Lagrangian mesh. For the convective energy transfer we use the time-dependent Kuhfu\ss{} (\cite{ku86}) convection model reformulated for the use in stellar pulsation codes (see also Wuchterl \& Feuchtinger \cite{wf98}). Radiation is described in the diffusion approximation. Below we provide a short summary of the model. For extensive description and details of numerical implementation we refer the reader to Smolec \& Moskalik (\cite{sm08a}).

The momentum, internal energy, and turbulent energy equations are
\begin{equation} \frac{{\rm d}u}{{\rm d}t}=-\frac{1}{\rho}\frac{\partial}{\partial r}\big(p+p_{\rm t}\big)+U_{\rm q}-\frac{GM_r}{r^2},\label{motion}\end{equation}
\begin{equation} \frac{{\rm d}E}{{\rm d}t}+p\frac{{\rm d}V}{{\rm d}t}=-\frac{1}{\rho}\frac{\partial\big[r^2\big(F_{\rm r}+F_{\rm c}\big)\big]}{r^2\partial r}-C,\label{energy}\end{equation}
\begin{equation} \frac{{\rm d}e_{\rm t}}{{\rm d}t}+p_{\rm t}\frac{{\rm d}V}{{\rm d}t}=-\frac{1}{\rho}\frac{\partial\big(r^2 F_{\rm t}\big)}{r^2\partial r}+E_{\rm q}+C.\label{turbulentenergy}\end{equation}
Above,  $u$ is the fluid velocity, which is a time derivative of the radius $u={\rm d}r/{\rm d} t$. $M_r$ is the mass enclosed in the radius $r$, $V$ is the specific volume which is the inverse of the specific density, $V=1/\rho$. $p$ and $E$ are the pressure and energy of the gas. $F_{\rm r}$, $F_{\rm c}$ and $F_{\rm t}$ are the radiative, convective, and turbulent fluxes, respectively. The radiative flux is computed assuming diffusion approximation. Radiation pressure and radiation energy are included in $p$ and $E$. The turbulent energy, $e_{\rm t}$, is computed according to the one-equation model of Kuhfu\ss{} (Eq.~\ref{turbulentenergy}). Turbulent energy equation and internal energy equation are coupled through the term $C$ of the form
\begin{equation}C=S-D-D_{\rm r},\end{equation}
where,
\begin{equation}S=\alpha\alpha_{\rm s}\frac{TpQ}{H_p} e_{\rm t}^{1/2}Y,\label{S}\end{equation}
\begin{equation}D=\alpha_{\rm d}\frac{e_{\rm t}^{3/2}}{\alpha H_p},\end{equation}
\begin{equation}D_{\rm r}=\frac{4\sigma\gamma_{\rm r}^2}{\alpha^2}\frac{T^3V^2}{c_p\kappa H_p^2}e_{\rm t}.\end{equation}
Above, $T$ is the temperature, $\sigma$ is the Stefan-Boltzmann constant, $Q=(\partial V/\partial T)_p$, $H_p$ is the pressure scale height, $c_p$ is the specific heat at a constant pressure and $\kappa$ is the opacity. The source (or driving) function, $S$, describes the rate of turbulent energy generation/damping through the buoyant forces. The source function is proportional to the superadiabatic gradient (dimensionless entropy gradient), $Y$,
\begin{equation}Y=\nabla-\nabla_{\rm a}=-\frac{H_p}{c_p}\frac{\partial s}{\partial r},\end{equation}
and therefore drives the turbulent energies in convectively unstable regions ($S>0$), and brakes the turbulent motions in convectively stable regions ($S<0$). Term $D$ models the decay of turbulent energy through the turbulent cascade. $D_{\rm r}$ describes the rate at which the turbulent energy is transformed into the internal energy through the radiative cooling of the eddies (see Wuchterl \& Feuchtinger \cite{wf98}). For the turbulent fluxes we have
\begin{equation} F_{\rm c}=\alpha\alpha_{\rm c}\rho T c_pe_{\rm t}^{1/2}Y,\label{Fc}\end{equation}
\begin{equation} F_{\rm t}=-\alpha\alpha_{\rm t}\rho H_p e_{\rm t}^{1/2}\frac{\partial e_{\rm t}}{\partial r},\end{equation}
\begin{equation} p_{\rm t}=\alpha_{\rm p}\rho e_{\rm t},\end{equation}
\begin{equation} U_{\rm q}=\frac{1}{\rho r^3}\frac{\partial}{\partial r}\bigg[\frac{4}{3}\alpha\alpha_{\rm m}\rho H_p e_{\rm t}^{1/2}r^3\bigg(\frac{\partial u}{\partial r}-\frac{u}{r}\bigg)\bigg],\end{equation}
\begin{equation} E_{\rm q}=\frac{4}{3}\alpha\alpha_{\rm m} H_p e_{\rm t}^{1/2}\bigg(\frac{\partial u}{\partial r}-\frac{u}{r}\bigg)^2.\end{equation}
Above, $p_{\rm t}$ is the turbulent pressure and $U_{\rm q}$ and $E_{\rm q}$ are the viscous momentum and energy transfer rates. We note that the turbulent viscosity always contributes to the driving of the turbulent energy, at the cost of pulsation.

The model equations contain 8 orders of unity scaling parameters,
mixing-length, $\alpha$ and parameters multiplying the turbulent
fluxes and driving/damping terms, $\alpha_{\rm p}$, $\alpha_{\rm
m}$, $\alpha_{\rm c}$, $\alpha_{\rm t}$, $\alpha_{\rm s}$,
$\alpha_{\rm d}$ and $\gamma_{\rm r}$. The theory provides no guidance
for their values, but some standard values are in use. These
values result from a comparison of the static, time-independent version of
the model with the standard mixing-length theory (see Wuchterl \&
Feuchtinger \cite{wf98} and Smolec \& Moskalik \cite{sm08a}). In practice, the values of the parameters
should be determined in a way that the models satisfy as many observational
constraints as possible. In principle, direct numerical simulations of convective zones can be used to calibrate the convective parameters, but currently no such simulations are available for large amplitude giant pulsators. The values of the parameters used in the
present paper are given in Table~\ref{convpar}. We note that set R3
is equivalent to set B of Baranowski et al. (\cite{bea09}) and was
adopted by these authors in successful modelling of the overall
properties of the radial velocity curves of first overtone Galactic
Cepheids.

\begin{table}
\caption{Three sets of convective parameters considered in this paper. Parameters $\alpha_{\rm s}$, $\alpha_{\rm c}$, $\alpha_{\rm d}$, $\alpha_{\rm p}$ and $\gamma_{\rm r}$ are given in the units of standard values ($\alpha_{\rm s}=\alpha_{\rm c}=1/2\sqrt{2/3}$, $\alpha_{\rm d}=8/3\sqrt{2/3}$, $\alpha_{\rm p}=2/3$ and $\gamma_{\rm r}=2\sqrt{3}$; see text below and Smolec \& Moskalik \cite{sm08a} for details).}
\label{convpar}
\centering
\begin{tabular}{ccccccccc}
\hline
Set & $\alpha$ & $\alpha_{\rm m}$ & $\alpha_{\rm s}$ & $\alpha_{\rm c}$ & $\alpha_{\rm d}$ & $\alpha_{\rm p}$ & $\alpha_{\rm t}$ & $\gamma_{\rm r}$\\
\hline
R1    & 1.5 & 0.30 & 1.0 & 1.0 & 1.0 & 1.0 & 0.0 & 0.75\\
R2    & 1.5 & 0.30 & 1.0 & 1.0 & 1.0 & 0.5 & 0.0 & 1.00\\
R3    & 1.5 & 0.50 & 1.0 & 1.0 & 1.0 & 0.0 & 0.0 & 1.00\\ %set B of Baranowski et al (2009)
\hline
\end{tabular}
\end{table}

%-_-_-_-_-_-_-_-_-_-_-_-_-_-_-_-_-_-_-_-_-_-_-_-_-_-_-_-_-_-_-_-_-_-_-_-_-_

\subsection{Importance of negative buoyancy\label{M.negativebuoyancy}}

In this section we look in detail at the important difference between the convection model adopted in our code and the convection model adopted in the Florida-Budapest hydrocode (e.g. Koll\'ath et al. \cite{koea02}). The latter code was used to compute all the convective double-mode Cepheid models published so far. As analysed by Smolec \& Moskalik (\cite{sm08b}), these double-mode models are physically incorrect, because they result from the exclusion of the
buoyant forces in convectively stable regions. In the Florida-Budapest hydrocode it is set
\begin{equation} S\propto Y_+,\label{SFB}\end{equation}
for the turbulent source function, and
\begin{equation} F_{\rm c}\propto Y_+, \end{equation}
for the convective heat flux, to be compared with $S\!\propto\!Y$
and $F_{\rm c}\propto Y$ in our code (Eqs.~\ref{S} and \ref{Fc}).
To exclude the buoyant forces in convectively stable regions, a
consequence of Eq. (\ref{SFB}) in the Florida-Budapest model, is incorrect and cannot be
justified. In convectively unstable layers ($Y>0$) the buoyant force is
responsible for the acceleration of turbulent eddies. In
convectively stable regions ($Y<0$), the buoyant force does not
disappear. It decelerates the turbulent eddies, restoring the
convective stability. Thus, we stress that the treatment of the source
function in the convectively stable regions, $S\propto Y$ in our code
and $S\propto Y_+$ in the Florida-Budapest code, is not arbitrary.
Even more importantly, the use of two different equations for the
source function (Eq. \ref{S} or \ref{SFB}) leads to significant
differences in the computed models. Particularly, the modal
selection is very different. As Smolec \& Moskalik (\cite{sm08b})
showed, when negative buoyancy is neglected (Eq.~\ref{SFB})
the non-resonant F+1O double-mode models can be easily computed. But the
mechanism responsible for the double-mode pulsation is artificial.
Because the negative buoyancy is excluded, the turbulent eddies cannot be
braked effectively below the envelope convection zone. Consequently,
a large region extending over several pressure scale heights
below the convection zone with relatively high turbulent energies
is present in these models. In this region eddy-viscous damping
strongly limits the amplitude of the fundamental mode; much stronger 
than it limits the amplitude of the first overtone. With its reduced
amplitude, the fundamental mode is no longer able to saturate the
pulsation instability alone, which allows the growth of the first
overtone and leads to stable double-mode pulsation. For more
details the reader is referred to Smolec \& Moskalik (\cite{sm08b})
and Smolec (\cite{rs09b}).

Our extensive survey (Smolec \& Moskalik \cite{sm08b}) in
search for stable F+1O double-mode pulsation in the physically
correct models, i.e. with negative buoyancy included, yielded
no result. We could not find a non-resonant double-mode model. However,
we have found some interesting F+1O resonant double-periodic models,
which we describe in Sect.~\ref{R.par}. We also performed the
model survey intended to reproduce the observed double-overtone
Cepheids in the LMC. Our results, also including double-overtone
models, are described in Sect.~\ref{R.LMC}.

%-_-_-_-_-_-_-_-_-_-_-_-_-_-_-_-_-_-_-_-_-_-_-_-_-_-_-_-_-_-_-_-_-_-_-_-_-_

\subsection{Mode selection analysis\label{M.modeselection}}

To search for stable double-periodic pulsation we adopt the methods developed by the Florida-Budapest group (e.g. Koll\'ath et al. \cite{koea02}). They are based on the analysis of results of direct hydrodynamical model integration with an amplitude equation formalism (e.g. Buchler \& Goupil \cite{bug84}). Here we provide a brief summary of the method. For details of our implementation the reader is referred to Smolec \& Moskalik (\cite{sm08b}).

\begin{figure*}
\centering
\resizebox{\hsize}{!}{\includegraphics{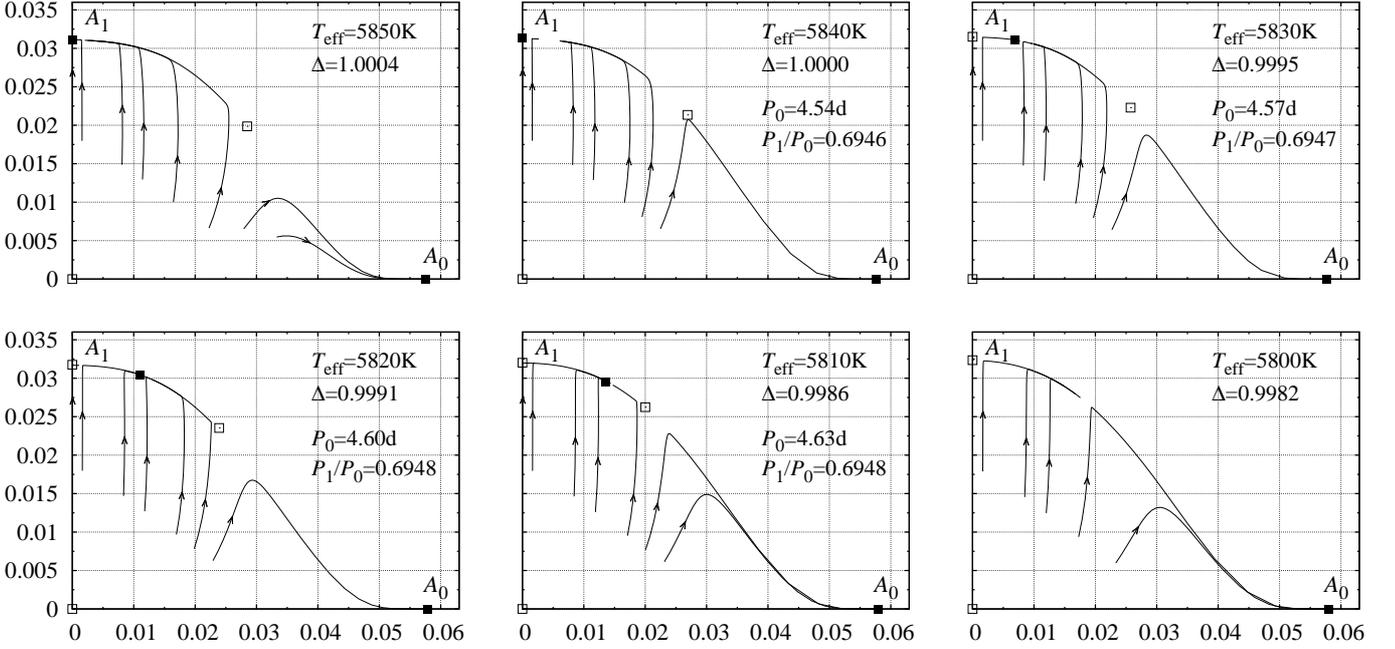}}
\caption{Fractional radius amplitude for the first
overtone ($A_1$) and for the fundamental mode ($A_0$) for six
consecutive Cepheid models of set R2. The models were initialized with a
range of amplitude ratios (bottom left parts of the diagrams) and
for each initialization the time evolution (marked with arrow) was
followed. The model properties are given in each panel. The proximity to the
resonance centre is characterised by
$\Delta=2\omega_1/(\omega_0+\omega_2)$. Solid and open squares mark
the location of stable and unstable fixed points computed through
fitting the hydrodynamical trajectories with non-resonant amplitude
equations. In panels three to five some trajectories evolve to a
double-periodic attractor with finite amplitude in both the fundamental
and first overtone modes. The remaining (rightmost) trajectories
calculated for these three models evolve towards a single-periodic
finite amplitude fundamental mode pulsation. }
\label{PARtraj}
\end{figure*}

The static model is initialized (or "kicked") with a scaled
mixture of linear velocity eigenvectors of the studied modes. Then,
a time-evolution of mode amplitudes follows using the analytical
signal method (see e.g. Koll\'ath et al. \cite{koea02}). The model
integration is stopped when full amplitude single-periodic
pulsation (limit cycle) is reached, or when the time evolution becomes
prohibitively slow, which may indicate the approach to the
attractor. The resulting trajectory is plotted in the
amplitude--amplitude diagram, similar to those shown in Fig.~\ref{PARtraj}. In this figure we plot on the vertical axis the fractional radius amplitude, $\delta R /R$, of the first overtone, and on the horizontal axis we plot the fractional radius amplitude of the fundamental mode. Each trajectory is represented by a single, continuous line starting in the bottom left part of the diagram, with the direction of the evolution marked by an arrow. For each model the non-linear integration
is repeated with several different initial conditions (range of amplitude ratios), chosen to
adequately explore the whole amplitude--amplitude phase-space.
The dimension of the phase-space can be large, depending on the
number of linearly excited modes and possible resonances. Linearly
damped modes that can be present in the initial phases of the model
integration (due to imperfect initialization) decay exponentially
and are of no interest, unless they are in resonance with other
linearly excited modes. In the latter case, their
amplitude can grow significantly.

Provided that the amplitude--amplitude phase-space is sufficiently
covered (several trajectories, sufficiently long integrations), the mode
selection can be deduced beyond doubt from the analysis of the
appropriate 2D amplitude--amplitude diagrams similar to those shown in
Figs.~\ref{PARtraj} and \ref{LMCtraj} (which are the sections
through the whole phase-space). For example, the trajectories presented
in first panel of Fig.~\ref{PARtraj} ($T_{\rm eff}=5850\,{\rm K}$)
evolve either towards fundamental mode single-periodic pulsation
(a fundamental mode limit cycle; two rightmost trajectories)
or towards first overtone single-periodic pulsation (six
leftmost trajectories, one of them overlaps with the vertical axis). In four consecutive panels of
Fig.~\ref{PARtraj}, the double-periodic attractor is clearly visible.
It is also evident that for these models the fundamental mode
limit cycle is stable, while the first overtone pulsation is unstable.
For the last model presented in Fig.~\ref{PARtraj} ($T_{\rm
eff}=5800\,{\rm K}$), all trajectories evolve towards fundamental
mode, which is the only stable pulsation state. More robust
conclusions can be drawn however if the hydrodynamic trajectories are
analysed with appropriate amplitude equations.

Amplitude equations can be used to compute the time-evolution of the mode amplitudes and phases. In particular, the mode selection for a given model can be established through computing the time-independent solutions of the amplitude equations (fixed points) and their stability. Stable fixed points are the attractors of the system and the trajectories evolve toward them. Unstable fixed points repel the trajectories. Fixed points correspond to steady non-linear pulsations. Stable, single-mode fixed points correspond to stable limit cycle pulsations.

In order to compute the fixed points and their stability, the linear
growth rates as well as non-linear self- and cross-saturation
coefficients of the modes have to be known. In a resonant case
it is also necessary to know the coupling coefficients of the interacting modes. These coefficients can be computed from the linear
eigenvectors and model structure, but this is too difficult, unless
strongly simplifying assumptions are made. In practice, saturation
and coupling coefficients can be derived through fitting the
appropriate amplitude equations to non-linear trajectories computed
with the hydrocode. The numerical procedure is straightforward in
a non-resonant case. The amplitude equations, which are complex in general, decouple
into a real part that yields the equations for the amplitudes of the
modes, and an imaginary part that yields the equations for their phases.
The equations for the amplitudes are used to derive the saturation
coefficients through a simple linear fit, while the equations for
the phases are not relevant in the non-resonant case.

In a resonant case, the situation is much more complicated. Full
complex amplitude equations have to be considered. The number of unknown
coefficients grows significantly, and a simple linear fit is no longer
possible. Therefore, the mode selection analysis, as described
above, is hard to conduct (see Smolec \cite{rs09b}). Below the modal 
selection for models in proximity of the
resonances is derived based on the analysis of hydrodynamic
trajectories only. As mentioned above and presented in the
forthcoming sections, the computation of several trajectories and 
sufficiently long integrations allows us to deduct the modal selection 
beyond doubt also in the resonant case.

%-_-_-_-_-_-_-_-_-_-_-_-_-_-_-_-_-_-_-_-_-_-_-_-_-_-_-_-_-_-_-_-_-_-_-_-_-_

\section{Resonant beat F+1O Galactic Cepheid models\label{R.par}}

There are only a few double-periodic Cepheids known in our Galaxy. The high extinction in the direction of the Galactic disc, where we expect to find these young objects, significantly limits our detection ability. Including very recent discoveries in all-sky surveys data, such as ASAS (Pojmanski \cite{po02}), there are 23 known F+1O double-periodic Cepheids (Antipin \cite{dm1,dm2,dm3}, Berdnikov \& Turner \cite{dm5}, Khruslov \cite{dm7}, Pardo \& Poretti \cite{dm11}, Wils et al. \cite{dm13}, Wils \& Otero \cite{dm14}). Only 15 1O+2O Cepheids were detected so far (Beltrame \& Poretti \cite{dm4}, Hajdu et al. \cite{dm6}, Khruslov \cite{dm8,dm9,dm10}, Pardo \& Poretti \cite{dm11}, Szczygiel \cite{dm12}), of which only two objects were known before 2009. In many cases the reddenings are unknown, making the derivation of physical parameters of these objects very uncertain. Here we focus on F+1O Cepheids. Observed period ratios, $P_1/P_0$, are confined between 0.697 and 0.713 for most of the objects. The typical periods of the fundamental mode are between two and six days (see Moskalik \& Ko\l{}aczkowski \cite{mk09}). The period ratio decreases with increasing period of the fundamental mode.

The double-periodic F+1O models presented in this section were found
accidentally during the model survey intended to reproduce the
Hertzsprung bump progression (Smolec \cite{rs09b}). A detailed linear
analysis revealed that the computed double-periodic behaviour can
be connected to the high-order parametric resonance,
$2\omega_1=\omega_0+\omega_2$. The same resonance was found to be
responsible for the double-periodic behaviour computed in radiative
$\beta$~Cephei models by Smolec \& Moskalik (\cite{sm07}).
$\beta$~Cephei stars are multi-periodic, mostly non-radial
pulsators, and two radial modes excited alone were so far not observed
in any star of this class. Consequently, the double-periodic
$\beta$~Cephei models of Smolec \& Moskalik (\cite{sm07}) are of
theoretical interest only. Yet they demonstrate that the
three-mode, high-order resonance can be conducive to producing
stable multi-mode pulsation. The convective $\delta$~Cephei models
described below confirm this finding and agree moreover fairly well
with the observations.

Static models were constructed using 150 mass zones, extending
down to $2.5\times 10^6\,{\rm K}$, with a fixed temperature ($T_{\rm
a}\!=\!11000\,{\rm K}$) in the anchor zone located 50 zones below
the surface. As discussed in Smolec \& Moskalik (\cite{sm08a}),
$T_{\rm a}\!=\!15000\,{\rm K}$ was used for model sequences
in which effects of turbulent pressure were turned on ($\alpha_{\rm
p}\!\ne\! 0$). The convective parameters of the model sequences that are
discussed in this section, R1, R2 and R3, are collected in
Table~\ref{convpar}.

The computed models are characterised by Galactic chemical composition, $X\!=\!0.7$, $Z\!=\!0.02$. OPAL opacities were used (Iglesias \& Rogers \cite{ir96}) and were supplemented at the low temperatures with the Alexander \& Ferguson (\cite{af94}) opacity data. Opacities were generated for the solar mixture of Grevesse \& Noels (\cite{gn93}). Models were computed along a sequence of constant mass and constant luminosity, and varying effective temperature. Masses of the models are $4.5\,{\rm M}_\odot$ for models adopting convective parameters of set R1, $4.75\,{\rm M}_\odot$ for models adopting convective parameters of set R2, and $5.0\,{\rm M}_\odot$ for models adopting convective parameters of set R3. Luminosities of the models were derived using the mass-luminosity relation resulting from the Schaller et al. (\cite{sch92}) evolutionary computations, $\log(L/{\rm L}_\odot)=3.56\log(M/{\rm M}_\odot)+0.79$.

\begin{figure}
\centering
\resizebox{\hsize}{!}{\includegraphics{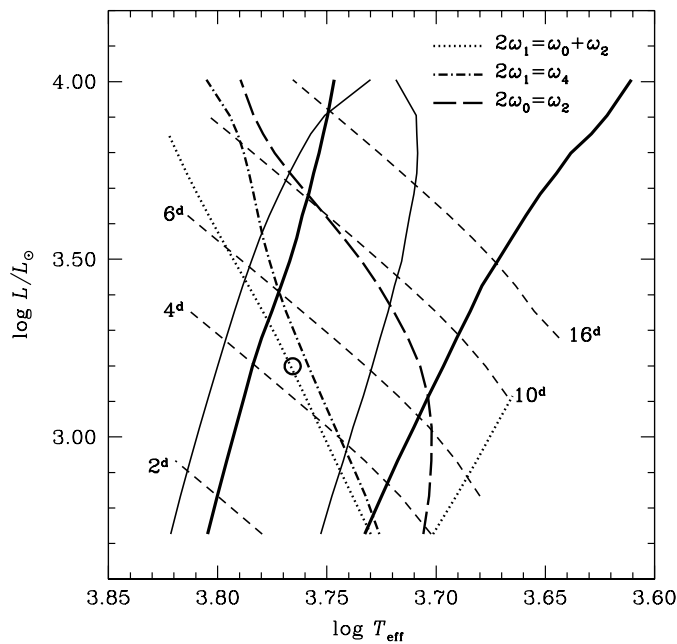}}
\caption{Location of the $2\omega_1=\omega_0+\omega_2$ resonance centre (dotted line) and the computed resonant multi-mode F+1O Cepheid models (within the circle) in the HR diagram. The loci of other resonance centres are also shown. The thick and thin solid lines enclose the fundamental and first overtone instability strips. The dashed lines are the lines of the constant period (F-mode) indicated in the figure. Computations were done for set R2 of Table~\ref{convpar}.}
\label{PARhr}
\end{figure}

The multi-mode solution was first found for one model of set R1. Detailed linear analysis revealed that the $2\omega_1=\omega_0+\omega_2$ resonance can be involved in the pulsations, because the model was located very close to the resonance centre. The proximity parameter, which we define for the resonance under discussion as
\begin{equation}
\Delta=\frac{2\omega_1}{\omega_0+\omega_2},\label{PAR.proximity}
\end{equation}
was equal to $\Delta=1.0006$. This finding motivated the search for
double-periodic F+1O Cepheid pulsation connected with the discussed
three-mode resonance. Multi-mode models were found also for sets R2
and R3 of Table~\ref{convpar}. The widest domain of
double-periodic solutions exists for parameter set R2. In
Fig.~\ref{PARtraj} we present the results of hydrodynamical model integrations
for six consecutive models of this set. In
Fig.~\ref{PARhr}, the loci of the $2\omega_1=\omega_0+\omega_2$
resonance in the HR diagram are plotted. Double-periodic models are
located within the circle, at $\log(L/{\rm L}_\odot)\approx 3.2$.
In Fig.~\ref{PARhr} we also plot the lines representing the loci
of other resonances, which are important for Cepheid pulsation,
$2\omega_0=\omega_2$ and $2\omega_1=\omega_4$.

\begin{table}
\caption{Properties of the computed double-periodic F+1O convective Cepheid models (mass, effective temperature, proximity parameter, fundamental mode period and $P_1/P_0$ period ratio). Convective parameters of the models are given in Table~\ref{convpar}.}
\label{PARprop}
\centering
\begin{tabular}{c|ccccc}

Set &  $M$[M$_\odot$]&$T_{\rm eff}$[K] & $\Delta$ & $P_0$[d] & $P_1/P_0$\\
\hline
R1 & 4.50 &5755 & 1.00064 & 4.21 & 0.6964\\
R2 & 4.75 &\multicolumn{4}{c}{see Fig.~\ref{PARtraj}}\\
R3 & 5.00 &5965 & 1.00009 & 4.77  & 0.6929\\
\hline
\end{tabular}
\end{table}

The mode selection scenario is evident from the hydrodynamical trajectories alone and is the same for all three parameter sets under discussion (see Fig.~\ref{PARtraj}). The double-periodic attractor always coexists with the stable fundamental mode attractor. The hysteresis domain (in which the mode selection depends on initial conditions) is always very narrow. Close to the resonance centre the consecutive models were computed in $5\,{\rm K}$ steps in the effective temperature. For sets R1 and R3 the double-periodic solution was found in only one model (see Table~\ref{PARprop} for its properties), indicating that the interesting domain is narrower than $10\,{\rm K}$. Only for models adopting convective parameters of set R2 the double-periodic domain is wider, and extends for more than $40\,{\rm K}$. For all three model sequences, this domain is located between the F/1O either-or domain (hotter models) and a fundamental-mode-only pulsation domain (cooler models). Therefore, from an astrophysical point of view, a double-periodic solution can be reached only during red-ward evolution by models previously pulsating in the first overtone. During the blue-ward evolution, models pulsate in the fundamental mode, which remains stable. Double-periodic pulsation is not possible.

The properties of the computed double-periodic models, their periods and
period ratios are collected in Table~\ref{PARprop} (sets R1 and R3)
and Fig.~\ref{PARtraj} (set R2). The models occupy a narrow period
range. Compared to the observations (see Petersen diagram in e.g.
Moskalik \& Ko\l{}aczkowski \cite{mk09}), the period ratios are slightly
too low. This can be easily compensated by the
decrease of model's metallicity. However, the main drawback of most
of the computed models is their excessive amplitude. Only for
models of set R3 do the amplitudes of the single-mode pulsation agree with
the observations. What is more important, the overall properties of the 
fundamental mode and first overtone Cepheids are reproduced reasonably well 
with the convective parameters of set R3 (see Smolec
\cite{rs09b} and Baranowski et al. \cite{bea09}, respectively).
For the models of sets R1 and R2, the amplitudes of the single-mode
solutions along the computed model sequences are always higher
than observed. Particularly the amplitude of the fundamental
mode is too high by up to 30 per cent. We will return to the
problem of excessive amplitudes later in this section.

At first glance, the non-resonant mechanism seems to be the most
likely cause of the double-periodic pulsation. Double-periodic
solutions are always located between the fundamental only and F/1O
either-or domains, which is natural for the non-resonant
mechanism. For the typical resonant excitation, the double-periodic
domain usually emerges in the middle of the single-mode pulsation
domain -- fundamental mode domain, as for the 2:1 resonance
discussed by Smolec (\cite{rs09a}), or first overtone domain, as for
the three-mode resonance acting in $\beta$~Cephei models (Smolec \&
Moskalik \cite{sm07}). This is because of the resonant destabilisation
of one of the limit cycles. Also, the non-resonant amplitude
equations seem to capture our hydrodynamical results very
well. The solid and open squares in Fig.~\ref{PARtraj} correspond to
stable and unstable fixed points, computed as described in
Sect.~\ref{M.modeselection}, through fitting the non-resonant
amplitude equations to the hydrodynamical trajectories. Except for one
model with $T_{\rm eff}=5840\,{\rm K}$, the location and stability
of the fixed points is consistent with the properties of the
computed trajectories.

Nevertheless, other arguments point towards a resonant explanation of
the computed double-periodic models. First, all the computed
double-periodic models are very close to the resonance centre.
The proximity parameter, $\Delta$ (Eq.~\ref{PAR.proximity}),
is given in Table~\ref{PARprop} for the two relevant models of sets
R1 and R3 and in Fig.~\ref{PARtraj} for the models of set R2. This is
not likely to happen by accident, particularly for the three model
sequences of the different convective parameters and the masses and
luminosities. Second, our extensive model survey (Smolec \&
Moskalik \cite{sm08b}) has not revealed any double-periodic
solutions; particularly, no such solutions have been found close to
the transition line between fundamental mode pulsation domain and
the F/1O either-or domain. We verified that in all models the
$2\omega_1=\omega_0+\omega_2$ resonance was far from the transition
line. Also, the double-periodic models were not found close to the
resonance centre if the transition line was distant from the resonance
centre. For example, models adopting the convective parameters of set R3
and slightly lower masses ($4.75\,{\rm M}_\odot$ instead of
$5.0\,{\rm M}_\odot$) all pulsate in the fundamental mode when they are 
computed close to the resonance centre. This indicates that a special
condition is required to generate the stable double-periodic
pulsation -- a proximity of the models to the resonance centre and
a proximity to the discussed transition line. Consequently,
the double-periodic domain is always restricted to a very narrow region
in the HR diagram, both in effective temperature and in luminosity.
At the moment we cannot provide any explanation why these two
conditions have to be satisfied simultaneously.

The existence and properties of the double-periodic pulsation depend on
the values of convective parameters that enter the model computations.
We computed several sequences of models with other convective
parameters than R1--R3. In order to decrease the model
amplitudes, which are too high for sets R1 and
R2, we computed some models with an increased eddy-viscous dissipation.
In particular, we computed a model sequence adopting the
convective parameters of set R1, except for the eddy-viscosity
parameter, $\alpha_{\rm m}$, which was set to a slightly higher value
($\alpha_{\rm m}=0.35$, instead of $\alpha_{\rm m}=0.30$).
Unfortunately, the double-periodic solution simply disappeared. However,
we note that it is hard to analyse how the existence and properties
of possible double-periodic models depend on their convective and
physical parameters. As noted above, the double-periodic models are
located close to the resonance centre and close to the transition
line between the F and F/1O pulsation domains. The change in e.g.
convective parameters while keeping the physical parameters of the
model fixed changes the relative location of these lines.
Consequently, the lack of the double-periodic solutions in models
computed close to the resonance centre may be caused not by the
'bad' convective parameters, but because we are
investigating a wrong part of the HR diagram. To
check whether the inclusion of radiative losses is crucial for the
existence of double-periodic solutions (we used $\gamma_{\rm
r}\ne 0$ in all three sets with double-periodic solutions), we
computed two additional model sequences, where all parameters were left exactly
the same as for sets R1 and R2, except for the radiative losses, which
were turned off ($\gamma_{\rm r}=0$). No double-periodic models were
found close to the resonance centre, but the transition line
between the F and F/1O pulsation domains also shifted towards higher
temperatures. Therefore, we cannot say whether the inclusion of
radiative losses is necessary or not. These definite claims
require a much more extensive model survey. Such a survey
was not conducted because the double-periodic resonant
solutions we found can model only a very small subgroup of the
observed double-periodic F+1O Cepheids and offers no general solution
to the reopened problem of modelling the beat Cepheid pulsation.

Nevertheless, the presented double-periodic models are promising,
because they are computed with a convective hydrocode that includes negative
buoyancy effects, and despite the described difficulties, their
parameters agree fairly well with the observations.

%-_-_-_-_-_-_-_-_-_-_-_-_-_-_-_-_-_-_-_-_-_-_-_-_-_-_-_-_-_-_-_-_-_-_-_-_-_

\section{LMC double-overtone Cepheid model survey\label{R.LMC}}

In contrast to our Galaxy, many more double-periodic Cepheids are known in the Large Magellanic Cloud. Thanks to observational projects aimed at detecting gravitational lensing events, such as MACHO (e.g. Alcock et al. \cite{al95}) or OGLE (e.g. Udalski et al. \cite{ogle}), a precise and extended photometry of many fields covering significant part of the LMC is available. The detected samples of Cepheids are large and homogeneous. Here we focus on data from the third phase of the OGLE project (Soszy\'nski et al. \cite{so08b}).

Double-overtone Cepheids are the most frequent class of
multi-periodic Cepheids in the LMC. Two hundred and six such variables were found in
OGLE-III data (Soszy\'nski et al. \cite{so08b}) compared with only 61 F+1O double-periodic
Cepheids. There are many more single-periodic first overtone
Cepheids (1238), but in a period range $0.5\thinspace{}{\rm
d}<P_1<0.9\thinspace{}{\rm d}$, double-overtone Cepheids are the most
common type of pulsators among the LMC Cepheids. Surprisingly, only
14 variables are confirmed as single-periodic second overtone
pulsators. Hence, the second overtone is excited predominantly together
with the first overtone.

The Petersen diagram for the LMC double-overtone Cepheids is plotted in
Fig.~\ref{LMCmodels}. The double-overtone Cepheids seem to fall into
two overlapping groups in this diagram. The first, less numerous
group extends from the shortest periods up to periods
slightly longer than $P_1 \approx 0.6\,$days ($\log P_1\approx
-0.2$). Within this group, one can observe a weak increasing trend
in the $P_2/P_1$ period ratio. Members of the second, more numerous
group that extends for longer periods ($P_1>0.6\,$days) form a well
defined sequence with a decreasing period ratio.

The evolutionary state of these objects is a matter of controversy. An analysis of the Petersen and period-luminosity diagrams (Dziembowski \& Smolec \cite{ds09a}) shows that short period objects are in the post-main sequence evolutionary phase, crossing the instability strip for the first time. However, objects with a longer period represent a challenge for the stellar evolution theory, because the inferred masses are too small for entering the instability strip during the core-helium burning, and the luminosities are much higher than in the post-main sequence phase (see Dziembowski \& Smolec \cite{ds09a}, and discussion in Sect.~\ref{sec.lmcle}).

A systematic survey in search for stable double-overtone Cepheid
pulsation was not published up to date. Some 1O+2O models were
computed with the Florida-Budapest hydrocode (see  Buchler
\cite{bu09}), but details of these models were not published.
Also, the exclusion of the negative buoyancy in the Florida-Budapest code
(Sect.~\ref{M.negativebuoyancy}) may influence the computed
double-overtone behaviour, just as for the F+1O double-mode
Cepheid models (Smolec \& Moskalik \cite{sm08b}). Therefore, it is
very important to check whether the double-overtone models can be
computed with the hydrocode, which includes negative buoyancy
effects. This survey is complementary to our (unsuccessful) search
for F+1O double-periodic Cepheids (Smolec \& Moskalik \cite{sm08b}),
and allows for more definite statements about the ability to model the
beat Cepheid pulsation with current convection models. The release
of OGLE-III data, which contain a significant number of double-overtone
Cepheids (Soszy\'nski et al. \cite{so08b}) provided additional
motivation, and focused our attention on modelling the LMC
pulsators.

\subsection{Construction of models}

The modelling of double-overtone pulsation and, more generally,
pulsation involving higher order overtones, is not an easy task.
The higher the order of the pulsation mode, the deeper into the envelope
it penetrates. Hence, to obtain reliable periods and period ratios,
model envelopes should be deep and computed with higher resolution
in the internal layers compared to fundamental mode Cepheid
models. On the other hand, double-overtone Cepheids are low mass
objects, characterised by small growth rates. This makes non-linear
computations extremely time-consuming and therefore a relatively coarse
mesh is necessary to conduct an extended non-linear model survey. In
order to get reliable periods and period ratios with a relatively
coarse mesh, the model structure, zoning, and depth of the envelope
in our code were chosen to reproduce the results obtained with the
LNA code of Dziembowski (\cite{wd77}) as closely as possible.
Dziembowski's code, which is coupled with the Warsaw-New Jersey
stellar evolution code (see e.g. Pamyatnykh \cite{ap99}),
allows the computation of deep, evolutionary models with high
spatial resolution, providing accurate periods and period ratios
that can be used in asteroseismic modelling (e.g. Moskalik \&
Dziembowski \cite{md05}).
However, the code adopts frozen-in MLT convection, which is much simpler
than the time-dependent treatment used in our code.
Therefore, a comparison of the model periods and period ratios computed
with both codes was conducted either for purely radiative models, or
with convection in the frozen-in approximation in both of the codes.
We note that with appropriate convective parameters, the Kuhfu\ss{}
model used in our envelope code can be reduced to a
standard MLT (Wuchterl \& Feuchtinger \cite{wf98}). 
The resulting mesh structure for our
envelope models is the following: the models consist of $200$ mass
zones, of which $50$ outer zones have an equal mass down to the anchor
zone, in which the temperature is set to $T_{\rm a}=11000\,{\rm K}$.
The envelope extends down to temperatures of $8\times 10^6\,{\rm K}$.

In the final linear and non-linear model computations only one set of
convective parameters was used namely set R3 of Table~\ref{convpar}. 
The non-linear computation of double-overtone pulsation is
extremely time-consuming and prohibits a more detailed parameter
study. Therefore, the convective parameters should be carefully chosen,
and we believe that set R3 represents a good choice. We note that
with the convective parameters of set R3, we were able to 
successfully model the overall properties of the radial velocity curves of
the first overtone Cepheids (Baranowski et al. \cite{bea09}). With
these convective parameters, the models of the radial velocity curves of
the fundamental mode Cepheids also agree with the observations (Smolec
\cite{rs09b}). It is also important that set R3 includes the effects
of radiative cooling of the convective elements ($\gamma_{\rm r}\ne
0$). The inclusion of this effect was necessary to reproduce the long
periods of some of the first overtone pulsators (Baranowski et al.
\cite{bea09}). It was also claimed that the inclusion of radiative
cooling was necessary to obtain the 1O+2O non-linear double-mode models
with the Florida-Budapest code (Buchler \& Koll\'ath \cite{bk00}).
Finally, we would like to point out that the models to be discussed
are hot and convection is not expected to be very strong. Therefore,
the results should not be very sensitive to the exact values of the
convective parameters.

\subsection{Linear modelling}\label{sec.lmcle}

The results of the linear model survey presented in this section,
particularly the implications for stellar evolution theory, were
published by Dziembowski \& Smolec (\cite{ds09a}, DS09 in the
following). In the present paper, the linear models provide a background
for non-linear pulsation modelling, and are briefly summarised below.

The pulsation models were constructed along evolutionary tracks computed
with the Warsaw-New Jersey stellar evolutionary code, which allows us 
to compute evolutionary phases before core helium ignition.
The computations were conducted for two values of metallicity
that are appropriate for the LMC, namely $Z=0.006$ and $Z=0.008$. For
the hydrogen abundance, $X=0.72$ was adopted. The OP opacities (Seaton
\cite{sea05}) and Asplund et al. (\cite{a04}) solar mixture were
used in the opacity computations. Rotation and overshooting from
the convective core were neglected. The mixing length parameter in the
evolutionary code was set to $\alpha_{\rm MLT}=1.5$.

In addition to pulsation models computed along evolutionary track ($\Delta\log L=0.0$ in the following), we have computed the models with artificially increased luminosity. We imposed higher $L$ values ($\Delta\log L=0.2$ or $\Delta\log L=0.4$) at the bottom boundary of our models, keeping their mass fixed. The computed models cover the whole instability strip. This luminosity increase was intended to model either (overlarge) overshooting during the main sequence evolution or the core helium burning phase (second and third crossings of the instability strip).

\begin{figure}
\centering
\resizebox{\hsize}{!}{\includegraphics{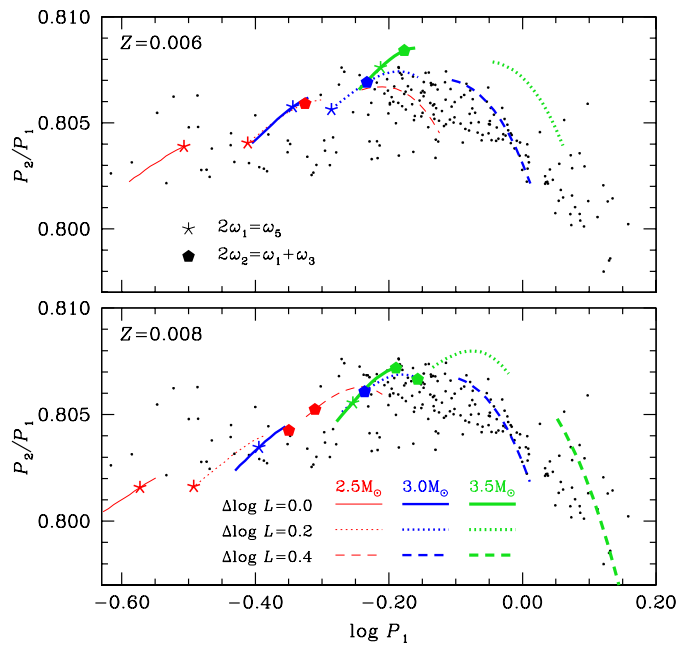}}
\caption{Domains of simultaneous linear instability of first and
second overtones in the Petersen diagram. Black dots represent
the observed LMC double-overtone Cepheids. Line segments are models
with normal and artificially increased luminosity, as described in
the bottom panel.  Asterisks and pentagons mark resonances between
low order modes. Resonances might be helping to  produce
double-overtone behaviour only for the shorter-period Cepheids, but
not for the more numerous longer-period ones.}
\label{LMCmodels}
\end{figure}

The domains of the simultaneous linear instability of the first and second
overtone modes are plotted in the Petersen diagrams presented in
Fig.~\ref{LMCmodels} (upper panel for $Z=0.006$ and lower panel
for $Z=0.008$). The corresponding period-luminosity (PL) diagrams were
published in DS09. The linear models well reproduce the observed
Petersen and PL diagrams. The inferred masses of the double-overtone
Cepheids are $3.0\pm0.5\,{\rm M}_\odot$. However, in order to
reproduce the objects of longer periods ($P_1>0.6\,{\rm days}$),
which represent the majority of the sample, a significant luminosity
increase compared to the post-main sequence evolutionary phase is
required, indicating that these objects are in the core helium burning
phase. As noted by DS09, this represents a challenge for the stellar
evolution theory. According to current evolutionary calculations,
the minimum mass of the star to enter the instability strip in its core
helium burning phase is higher than $4.0\,{\rm M}_\odot$ (see
references in DS09). For a discussion of this point we refer the
reader to DS09.

Our detailed linear analysis revealed that two types of resonances
can be conducive to producing stable double-overtone pulsation.
These are 2:1 resonance between the first overtone and linearly
damped fifth overtone, $2\omega_1=\omega_5$, and high-order
resonance, involving the three lowest order overtones,
$2\omega_2=\omega_1+\omega_3$. The loci of resonance centres within
the 1O+2O instability domains (and slightly beyond) are shown with
asterisks ($2\omega_1=\omega_5$) and pentagons
($2\omega_2=\omega_1+\omega_3$) in Fig.~\ref{LMCmodels}. The first
resonance, $2\omega_1=\omega_5$,  was suggested by Dziembowski (see
Soszy\'nski et al. \cite{so08b}, DS09) as a possible factor shaping
the structures visible in Fourier decomposition parameters of
the shortest period first overtone Cepheids. It is also well
known that 2:1 resonances can be conducive to producing stable
multi-mode pulsation, as shown theoretically by e.g. Dziembowski \&
Kov\'acs (\cite{dk84}). Examples of double-periodic hydrodynamical
radiative models of both RR~Lyrae and Cepheids in the proximity of
a 2:1 resonance are also known (Kov\'acs \& Buchler
\cite{kb88}, Buchler, Moskalik \& Kov\'acs \cite{bmk90}, Smolec
\cite{rs09a}). The latter resonance, $2\omega_2=\omega_1+\omega_3$,
is of the same type as the resonance operating in the F+1O
double-periodic Cepheid models described in Sect.~\ref{R.par}. We
note that these resonances can be operational only in objects
of periods shorter than $P_1<0.7\,{\rm days}$. At longer periods,
the non-resonant mechanism should be responsible for the observed
double-overtone pulsation.

\subsection{Non-linear modelling}

An extensive non-linear model survey was conducted for models with metallicities $Z=0.006$ and $Z=0.008$. Model sequences were computed along selected sequences described in the previous section chosen to cover the significant part of the Petersen diagram.

\begin{figure}
\centering
\resizebox{\hsize}{!}{\includegraphics{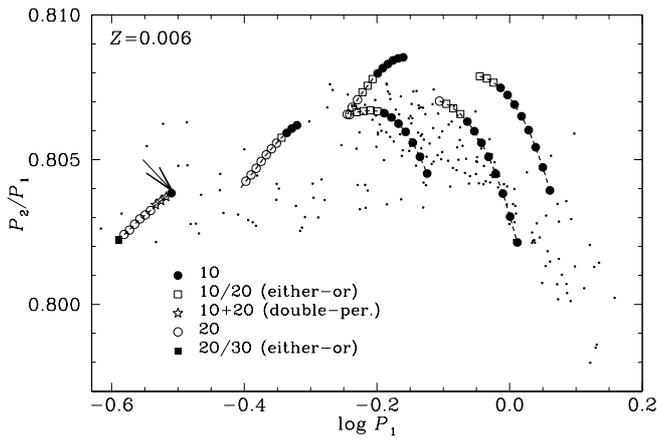}}
\caption{Petersen diagram showing the location and stability
information for the computed non-linear convective models. The
models were chosen to cover the significant part of the Petersen
diagram. These are models of $2.5\,{\rm M}_\odot$,  $3.0\,{\rm
M}_\odot$ and  $3.5\,{\rm M}_\odot$ located along evolutionary
tracks ($\Delta\log L=0.0$) and models with artificially increased
luminosity of masses $3.5\,{\rm M}_\odot$ ($\Delta\log L=0.2$), and
$2.5\,{\rm M}_\odot$ and  $3.0\,{\rm M}_\odot$ ($\Delta\log L=0.4$).
All computations are for the metallicity parameter $Z=0.006$. Black dots
represent the observed LMC double-overtone Cepheids. The only
double-periodic models are found near the $2\omega_1=\omega_5$
resonance centre, which is marked with an arrow.}
\label{LMCmsel6}
\end{figure}

For metallicity $Z=0.006$, six non-linear model sequences were
computed. Three sequences are located within the instability domains
along evolutionary tracks ($\Delta\log L=0.0$) for stars of masses
$2.5\,{\rm M}_\odot$,  $3.0\,{\rm M}_\odot$ and  $3.5\,{\rm
M}_\odot$ (solid lines in Fig.~\ref{LMCmodels}). The other
three are located along horizontal paths with artificially increased
luminosity ($2.5\,{\rm M}_\odot$, $\Delta\log L=0.4$;  $3.0\,{\rm
M}_\odot$, $\Delta\log L=0.4$;  $3.5\,{\rm M}_\odot$, $\Delta\log
L=0.2$). The location of all the computed models is shown in
Fig.~\ref{LMCmsel6}.

\begin{figure}
\centering
\resizebox{\hsize}{!}{\includegraphics{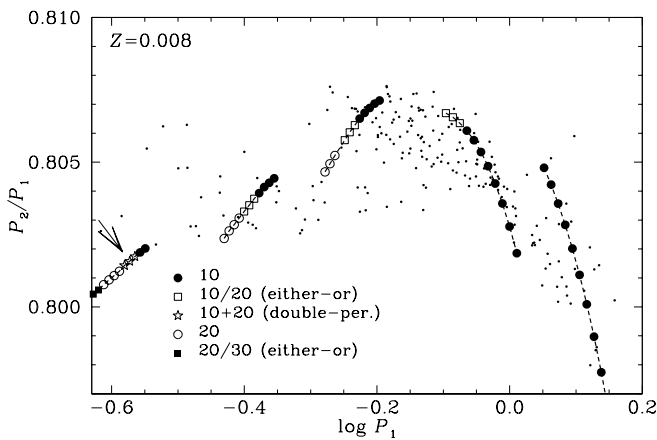}}
\caption{Same as Fig.~\ref{LMCmsel6}, but for metallicity
parameter $Z=0.008$. The chosen models cover the significant part of
the instability strip. These are models of $2.5\,{\rm M}_\odot$,
$3.0\,{\rm M}_\odot$ and  $3.5\,{\rm M}_\odot$ located along
evolutionary tracks ($\Delta\log L=0.0$) and models with an artificially
increased luminosity of masses $2.5\,{\rm M}_\odot$ and  $3.0\,{\rm
M}_\odot$ ($\Delta\log L=0.4$). The only double-periodic models are
found near the $2\omega_1=\omega_5$ resonance centre, which is marked with an
arrow.} 
\label{LMCmsel8}
\end{figure}

Five non-linear model sequences were computed for a higher metallicity, $Z=0.008$. Three sequences correspond to first crossing evolutionary models of masses, $2.5\,{\rm M}_\odot$, $3.0\,{\rm M}_\odot$ and $3.5\,{\rm M}_\odot$. The remaining two sequences correspond to horizontal paths with an artificially increased luminosity ($3.0\,{\rm M}_\odot$, $\Delta\log L=0.4$ and $3.5\,{\rm M}_\odot$, $\Delta\log L=0.4$). The location of the computed models is shown in Fig.~\ref{LMCmsel8}.

\begin{figure*}
\centering
\resizebox{\hsize}{!}{\includegraphics{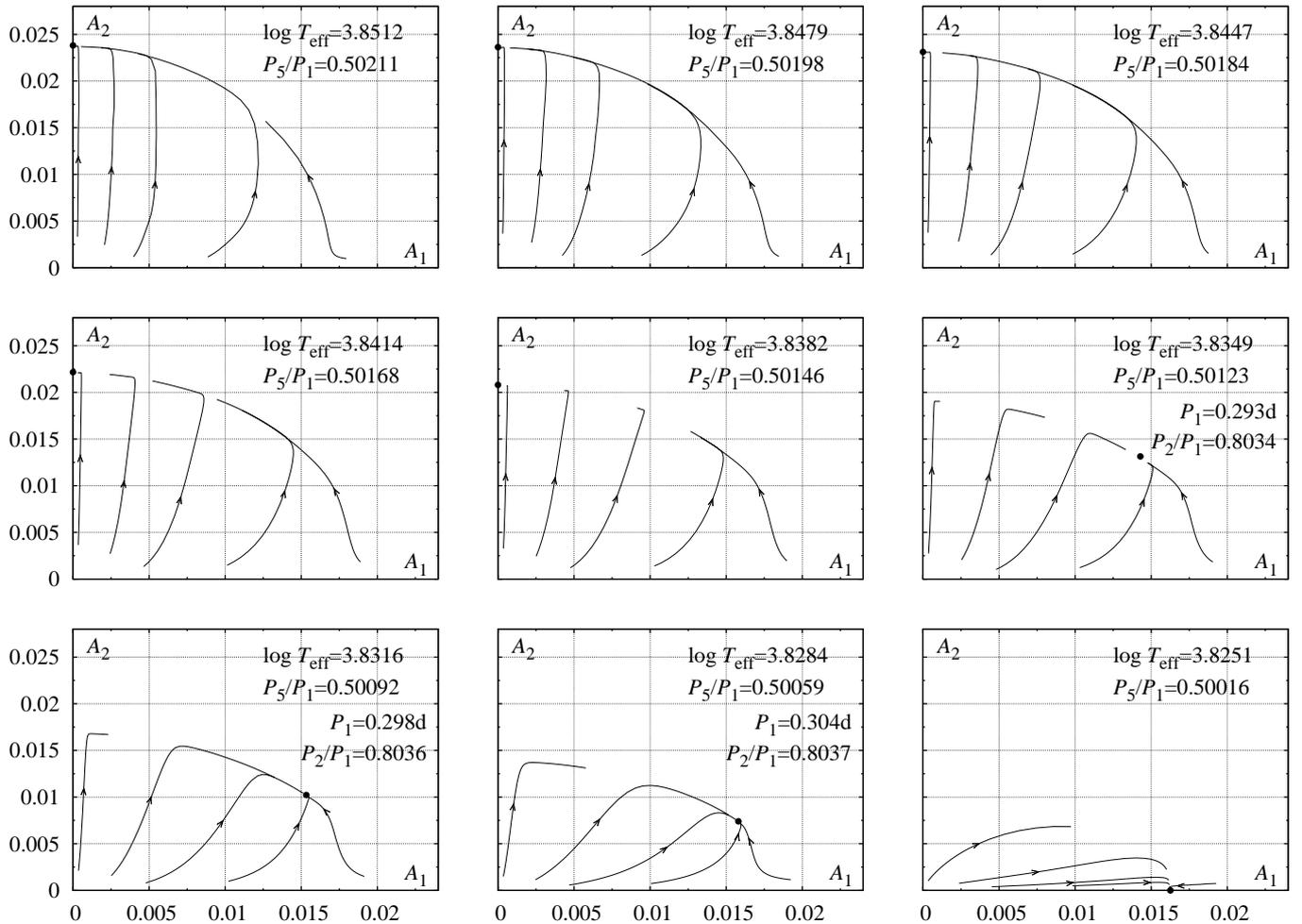}}
\caption{Fractional radius amplitude for the second
overtone ($A_2$) and for the first overtone mode ($A_1$), for nine
consecutive models located along evolutionary track of $2.5\,{\rm
M}_\odot$ ($Z=0.006$). The models were initialized with a range of
amplitude ratios. The effective temperatures of the models and
linear $P_5/P_1$ period ratios are given in each panel. The approximate 
location of the attractors is marked with filled circles. For the models
with a double-periodic attractor the periods $P_1$ and period ratios,
$P_2/P_1$, are also given. All trajectories in the first five panels
evolve towards a single-periodic second overtone attractor. In the next
three panels (panels 6--8) a double-periodic attractor is clearly
visible. In the last panel, all trajectories evolve towards a
single-periodic first overtone attractor.}
\label{LMCtraj}
\end{figure*}

For each model in a sequence five hydrodynamical integrations were
conducted, each initialized with a different mixture of the first and
second overtone's velocity eigenvectors. For most of the models
the integrations were carried over 8000 pulsation cycles. For the
least massive models located along the evolutionary track
($2.5\,{\rm M}_\odot$, $\Delta\log L=0.0$), integrations were twice
as long, because the growth rates are very low for these models. Some
individual trajectories for models located along the evolutionary track
with $2.5\,{\rm M}_\odot$ ($Z=0.006$) are presented in
Fig.~\ref{LMCtraj}.

The modal selection information for the computed models is presented
with different symbols in Figs.~\ref{LMCmsel6} and \ref{LMCmsel8}.
It was derived through the analysis of the computed trajectories
(presented e.g. in Fig.~\ref{LMCtraj}). As described in
Sect.~\ref{M.modeselection}, hydrodynamical computations were not
analysed with amplitude equations because of the
resonances. We also note that for the discussed models the fundamental
and third overtone modes can be linearly unstable as well.
We checked with the analytical signal method that these two modes are
present during the transient evolution, but they decay, and
except for one model (see next paragraph) are not present in full
amplitude pulsation.

The results are qualitatively the same for both metallicities, and below
we focus our discussion on the lower metallicity models
(Figs.~\ref{LMCmsel6} and \ref{LMCtraj}). The most interesting
modal selection scenario is visible along the $2.5\,{\rm M}_\odot$
evolutionary track. The individual trajectories of some models in this
sequence are displayed in Fig.~\ref{LMCtraj}. For the hottest
model in this sequence (not shown in Fig.~\ref{LMCtraj}), 2O/3O
hysteresis is possible. Depending on the initial conditions,
the trajectories evolve either towards the second overtone attractor or
towards the third overtone attractor. In six consecutive cooler
models, only single-periodic pulsation in the second overtone is
possible. Then, a double-overtone domain emerges. A multi-mode
attractor is clearly visible for models with $\log T_{\rm
eff}=3.8349,\ 3.8316,\ {\rm and}\ 3.8284$. Periods and period ratios
for these models are given in Fig.~\ref{LMCtraj}. The only model
redward of this domain pulsates in the first overtone. Thus,
the double-overtone domain appears in between the first overtone
and the second overtone pulsation domains. Although this is
a typical scenario for a non-resonant mechanism, a
resonant mechanism cannot be excluded {\it a priori}. The arrow in
Fig.~\ref{LMCmsel6} shows the location of the $2\omega_1=\omega_5$
resonance centre. It is located quite close to the double-overtone
domain. The resonant destabilisation of the first overtone limit
cycle, if it occurs at a proper position along the model
sequence, can in principle lead to the emergence of the
double-overtone domain instead of the either-or 1O/2O domain.
Note that the 2:1 resonance can affect the
pulsations in a very wide range of period ratios, as is observed
for bump Cepheids, and that destabilisation may not necessarily
occur exactly at the resonance centre (see Smolec \cite{rs09a}). We
postpone the discussion of the nature of the computed
double-periodic models to the next paragraphs in this section.

The remaining five non-linear model sequences of $Z=0.006$ all display qualitatively the same mode selection scenario, which is different however from scenario described for the $2.5\,{\rm M}_\odot$ sequence. A double-periodic domain is not present. Instead, the either-or 1O/2O domain is located in between the single-periodic first overtone and second overtone pulsation domains. The discussed resonances seem to have no visible effect on the modal selection along these sequences.

For higher metallicity models, the overall modal selection is very
similar to the lower metallicity models (compare
Figs.~\ref{LMCmsel6} and ~\ref{LMCmsel8}). Again, the most interesting
results are obtained for models of the shortest periods with masses
of $2.5\,{\rm M}_\odot$. Here, a multi-mode domain emerges.
Simultaneous pulsation in the first and second overtones is possible
for three models along the sequence. The centre of the 2:1 resonance
between first and fifth overtones, marked with arrow in
Fig.~\ref{LMCmsel8}, falls exactly in the middle of the
double-periodic domain. This supports the hypothesis that the
resonance is crucial in establishing the stable double-periodic behaviour
that we found for both metallicities. For the remaining four model
sequences with metallicity $Z=0.008$, no traces of multi-mode
behaviour are found. Again, as for the $Z=0.006$ models, it seems
that resonances do not affect the modal selection along model
sequences of higher masses (longer periods).

Considering the double-overtone models we found it is hard to
judge which mechanism, resonant or non-resonant, is responsible for
the computed beat pulsation. Both double-periodic domains (for
$Z=0.006$ and $Z=0.008$) are located very close to the
$2\omega_1=\omega_5$ resonance centre. For the $Z=0.008$ models
the resonance centre falls exactly in the middle of the
double-periodic domain. This suggests the resonant origin of the
computed beat pulsation. On the other hand, these double-periodic
domains are located in between two single-periodic pulsation
domains, first overtone pulsation domain (to the red) and second
overtone pulsation domain (to the blue). This is typical for the
non-resonant scenario. Also, for all studied sequences of
masses ${\rm M} > 2.5\,{\rm M}_\odot$, which cross the
$2\omega_1=\omega_5$ resonance centre (see Fig.~\ref{LMCmodels}),
no traces of beat pulsation were found. Instead, an 1O/2O either-or
domain is present. Although the presence of the resonance may be not
sufficient to excite the beat pulsation and other factors may
be necessary (like for the F+1O models discussed in the
previous section), it is hard to identify these factors and
prove their necessity. Therefore, we do not presume to know which
mechanism, resonant or non-resonant, underlies the double-overtone
behaviour we found.

Although some double-overtone models were found, the overall results are
not satisfactory. The computed double-overtone domains are narrow and
are located at $P_1\approx 0.3\,{\rm d}$, i.e. at the
short-period end of the observed domain of the double-overtone pulsation
in the LMC. No double-overtone models were found at longer periods,
particularly in a period range, $0.5\,{\rm d}<P_1<0.9\,{\rm d}$, where
the double-overtone pulsation is the most common form of pulsation
in the LMC. Surprisingly, in many models only pulsation in the
second overtone is possible. Observationally, this form of pulsation
is very rare, as only 14 such objects are identified in the
LMC (Soszy\'nski et al. \cite{so08b}, still,
observational selection can be a factor here). Therefore, also 
for the 1O+2O double-overtone Cepheids the convective hydrocode fails
to reproduce the observed modal selection. This conclusion is not as
strict as for the F+1O Cepheids, because only one set of
convective parameters was explored. On the other hand, for hot
overtone models, convection is not expected to play a crucial role,
and results should not depend strongly on the values of convective
parameters and/or effects included in the convective model.

%-_-_-_-_-_-_-_-_-_-_-_-_-_-_-_-_-_-_-_-_-_-_-_-_-_-_-_-_-_-_-_-_-_-_-_-_-_

\section{Conclusions\label{C}}

The longstanding problem of non-linear modelling of beat Cepheid
pulsation remains open. Our search for stable double-periodic
Cepheid pulsation with convective hydrocode, which correctly
includes negative buoyancy effects, yielded unsatisfactory results.
This search is a difficult and time-consuming task. The turbulent convection model we use contains several free parameters. Also, the Cepheid domain extends over a wide range of luminosities in the HR diagram, covering different masses and evolutionary stages. Consequently, extensive model computations are needed. Our recent model surveys (Smolec \& Moskalik \cite{sm08b}, Smolec \cite{rs09b}, this paper) cover different stellar systems (Galaxy, LMC) and different evolutionary stages (post-main sequence and helium burning objects). Different sets of convective parameters were investigated. An extensive parameter study was done for F+1O Cepheid models (Smolec \& Moskalik \cite{sm08b}). For the 1O+2O models presented here, only a limited study with carefully chosen parameters was possible, owing to the longer computation time needed for these models (see Sect.~4.1). Only a few double-periodic models, limited to a narrow period
ranges, were found. In particular, we have not found any large domain in which the non-resonant mechanism is the sole cause of the double-mode pulsation.

In most of the observed Cepheids a non-resonant mechanism should be
responsible for the simultaneous pulsation in two modes. With purely
radiative hydrocodes, non-resonant double-mode models are found only
for RR~Lyrae stars, and only if the artificial viscosity is significantly
reduced (Kov\'acs \& Buchler \cite{kb93}). These models are however
sensitive to numerical details and their pulsation amplitudes are
too high. On the other hand, several computed resonant radiative
beat Cepheid and RR~Lyrae models (Kov\'acs \& Buchler \cite{kb88},
Buchler, Moskalik \& Kov\'acs \cite{bmk90}, Smolec \cite{rs09a})
disagree with the observations. The inclusion of turbulent convection
into pulsation hydrocodes seemed to offer the solution. Using
modified Kuhfu\ss{} convection model where negative buoyancy
effects were excluded (e.g. Koll\'ath et al. \cite{koea98}), the
Florida-Budapest group published many non-resonant beat Cepheid
models (e.g. Koll\'ath et al. \cite{koea02}, Buchler \cite{bu09}).
However, the exclusion of the negative buoyancy is physically not
justified and, as shown by Smolec \& Moskalik (\cite{sm08b}), it is
the main cause of the computed double-mode pulsation. Using the original
Kuhfu\ss{} prescription including the negative buoyancy, we were
unable to find any non-resonant F+1O double-mode Cepheid models
(Smolec \& Moskalik \cite{sm08b}). In this paper we conducted
an additional search for double-periodic Cepheids, including
double-overtone (1O+2O) pulsators. With our hydrocode, which
includes negative buoyancy effects, we found several
double-periodic models of both F+1O and 1O+2O type. In the
first case, the double-periodic pulsation is most likely caused by the
$2\omega_1=\omega_0+\omega_2$ resonance. In the latter case, the
exact mechanism cannot be identified beyond doubt. We only note that
the 2:1 resonance, $2\omega_1=\omega_5$ may be operational in these
models. The computed models are very interesting. Their periods and
period ratios agree well with the observations. Nevertheless,
they are restricted to very narrow domains in the HR diagram and
offer no general solution to the problem of modelling the beat
Cepheid pulsation.

Turbulent convection is an important phenomenon determining
the properties of Cepheid envelopes and cannot be neglected in pulsation
models. However, the numerical expense of non-linear computations requires
relatively simple convective recipes. So far, only one-equation
models for the generation of turbulent convection were implemented in
pulsation hydrocodes. Two such models were used in the computation of
Cepheid models, the Kuhfu\ss{} (\cite{ku86}) model and the
Stellingwerf model (Stellingwerf \cite{st82}, Bono \& Stellingwerf
\cite{bos92}). Our extensive computations (Smolec \& Moskalik
\cite{sm08b}, Smolec \cite{rs09b}, this paper) indicate that the
original Kuhfu\ss{} model (i.e. including negative buoyancy) is
incapable of reproducing the majority of the observed beat
Cepheids. In the original Stellingwerf (\cite{st82}) model, the negative
buoyancy effects were neglected, because in that model the turbulent source
function is proportional to square-root of the superadiabatic gradient,
$S\propto\sqrt{Y}$. This makes the original Stellingwerf model very
similar to the modified Kuhfu\ss{} model used in the
Florida-Budapest hydrocode, and indeed, as investigated by Buchler
\& Koll\'ath (\cite{bk00}) both convection recipes lead to
qualitatively the same results. Stellingwerf's (\cite{st82})
functional form of the source function was criticised by Gehmeyr \&
Winkler (\cite{gw92}) who pointed out the arising problems
associated with the long time-scale of decay of the turbulent eddies
in regions that became convectively stable during pulsation. 
Bono \& Stellingwerf (\cite{bos92}) modified the original
Stellingwerf recipe through setting $S\propto{\rm
sgn}(Y)\sqrt{|Y|}$. With this model for negative buoyancy, plenty of
Cepheid models were computed with the Italian code (e.g. Bono,
Marconi \& Stellingwerf \cite{bms00}). However, no double-periodic
classical Cepheid models were ever found (Bono, private communication).
 This supports our results. Including the negative
buoyancy in the convective models (both in Kuhfu\ss{} and
in Stellingwerf models), wide domains of double-periodic pulsation
cannot be found. We conclude that with the one-equation models used
in non-linear Cepheid modelling, double-periodic Cepheid pulsation
cannot be modelled satisfactorily.

The question of how the hydrocodes should be modified to solve the
puzzle of beat Cepheid pulsation is difficult to answer. A simple
convection model seems the most severe shortcoming of the present
hydrocodes. More sophisticated convective recipes are being
developed (St\"okl \cite{stokl}, Buchler \cite{bu09}), however, many
problems have to be solved before they can be applied in non-linear
modelling. Another shortcoming of the present models is the simple
structure of the constructed Cepheid envelopes -- they are
chemically homogeneous and non-rotating. Without detailed model
computations however, it is hard to judge whether e.g. possible
differential rotation could have an effect on modal selection.
The treatment of radiation is also very simple in most of the
hydrocodes, i.e. diffusion approximation is used (our code, the
Florida-Budapest code and Italian code). However, a more
detailed time-dependent treatment did not lead to substantial changes
in the computed models (Feuchtinger, Buchler \& Koll\'ath
\cite{fbk00}). Finally, in the present models only radial
pulsation is considered. Non-radial modes are neglected. There is a
growing evidence that non-radial modes are excited in classical
Cepheids, including the beat Cepheids (see e.g. Moskalik \&
Ko\l{}aczkowski \cite{mk09}). The coupling between non-radial and
radial modes can affect the modal selection, and in our opinion,
this is one of the ideas that should be investigated
first. For this purpose, an analytical approach based
on the amplitude equation formalism may be used, without the
necessity of developing a non-linear non-radial hydrocode.

\begin{acknowledgements}
We are grateful to Giuseppe Bono for his comments concerning the computation of double-mode classical pulsator models with Italian code. Model computations presented in this paper were completed while RS was PhD student in the Copernicus Centre, Warsaw, Poland. RS is supported by the Austrian Science Fund (FWF project AP 2120521). Part of this work was supported by the Polish MNiSW Grant No. 1 P03D 011 30. The grant N N203 379636 to Alosha Pamyatnykh is acknowledged for supporting the accommodation costs during RS's stay in Warsaw.
\end{acknowledgements}

\end{document}